\documentclass[aps,prb,reprint,groupedaddress,longbibliography,nofootinbib]{revtex4-2}

\usepackage[T1]{fontenc}
\usepackage{mathtools,amssymb,slashed,bbm,tensor,verbatim}
\usepackage{tikz}
\usetikzlibrary{decorations.markings,decorations.pathreplacing,arrows,arrows.meta,calc,shapes.arrows,shapes.geometric}
\usepackage{array,multirow}

\DeclareRobustCommand{\d}{\mathrm{d}}

\DeclareRobustCommand{\i}{\mathrm{i}}

\newcommand{\adjvec}[1]
{
	\rotatebox[origin=c]{90}{$\begin{pmatrix} \rotatebox[origin=c]{-90}{$\begin{matrix} #1 \end{matrix}$} \end{pmatrix}$}
}
\usepackage{xprintlen}

\begin{document}

\title{Appearance of Odd-Frequency Superconductivity in a Relativistic Scenario}

\author{Patrick J. Wong}
\affiliation{Department of Physics, University of Connecticut, Storrs, Connecticut 06269, USA}
\affiliation{Nordita, Stockholm University and KTH Royal Institute of Technology, Hannes Alfv\'ens v\"ag 12, SE-106 91 Stockholm, Sweden}

\author{Alexander V. Balatsky}
\affiliation{Department of Physics, University of Connecticut, Storrs, Connecticut 06269, USA}
\affiliation{Nordita, Stockholm University and KTH Royal Institute of Technology, Hannes Alfv\'ens v\"ag 12, SE-106 91 Stockholm, Sweden}

\date{\today}

\begin{abstract}
Odd-frequency superconductivity is an exotic  superconducting state in which the symmetry of the gap function is odd in frequency. Here we show that an inherent odd-frequency mode emerges dynamically under application of a Lorentz transformation of the anomalous Green function with the general frequency-dependent gap function. To see this, we consider a Dirac model with quartic potential and perform a mean-field analysis to obtain a relativistic Bogoliubov--de~Gennes system. Solving the resulting Gor'kov equations yields expressions for relativistic normal and anomalous Green functions. The form of the relativistically invariant pairing term is chosen such that it reduces to BCS form in the non-relativistic limit. We choose an ansatz for the gap function in a particular frame which is even-frequency and analyze the effects on the anomalous Green function under a boost into a relativistic frame. The odd-frequency pairing emerges dynamically as a result of the boost. In the boosted frame the order parameter contains terms which are both even and odd in frequency. The relativistic correction to the anomalous Green function to first order in the boost parameter is completely odd in frequency. In this paper, we provide evidence that odd-frequency pairing may form intrinsically within relativistic superconductors.
\end{abstract}

\maketitle

\section{Introduction}

Berezinskii odd-frequency pairing~\cite{berezinskii,balatskyabrahams,oddfreview,tanakareview,tanakascjunc,blackschafferreview} is a class of pairing in Fermi systems in which the paired condensate is odd in relative time between the paired particles.
The parity of a paired state of fermions may be classified according to the discrete permutation symmetries of 
spin permutation $S$, 
relative coordinate permutation $P^*$, 
orbital permutation $O$, and
time permutation $T^*$. 
To satisfy the Pauli principle the symmetries of a Fermi pair must obey the condition of
$S P^* O T^* = -1$ \cite{oddfreview,tanakareview,tanakatopsc}.
With respect to these permutation symmetries, odd-frequency pairing possesses $T^* = -1$. The time permutation and coordinate permutation symmetry operations  $T^*, P^*$ are swap operations where relative indices are permuted yet no global inversion (for coordinates) nor the global time reversal is invoked: These permutations are distinct from time reversal and spatial inversion operations, hence we use $^*$ to mark the difference. Hereafter we consider the single band pairing states and hence drop $O$. We thus focus on the $S P^* T^* = -1$  product. Table~\ref{tab:Tab1} shows the allowed symmetry states.
\begin{table}[htp!]
    \centering
    \begin{tikzpicture}
    \node at (0,0) 
    {\begin{tabular}{|w{c}{2em}||w{c}{2em}|w{c}{2em}|w{c}{2em}|w{c}{2em}|}
    \hline
      $S^{\phantom{*}}$   &  $-$ & $-$ & $+$ & $+$ \\
      $P^*$ &  $+$ & $-$ & $-$ & $+$ \\
      $T^*$ &  $+$ & $-$ & $+$ & $-$ \\\hline
    \end{tabular}};
    \draw[-latex] (-0.75,-0.65) arc(10:180:-0.4 and -0.5);
    \end{tikzpicture}
    \caption{Permutation parities of spin-singlet and spin-triplet pairing states are listed. The induction of the odd-frequency state is possible once we allow the conversions of $P^* = +1$ into $P^* = -1$, as shown by the arrow.  Lorentz boost mixes up space and time indices  and induces finite amplitude of $P^* = -1$ states. Similar considerations can be done for spin triplet states, $S = +1$, which we list here for completeness.}
    \label{tab:Tab1}
\end{table}
The table implies that odd-frequency states are allowed by the symmetry classification and appear on equal footing with their even-frequency counterparts (BCS), so they therefore should in principle appear as ubiquitously as even-frequency states. We emphasize that the $T^*=-1$ states appear with equal weight as the $T^*=+1$ states in the table.
However, the majority of work on superconductivity, both theoretical and experimental, has focused only on even-frequency pairing, with odd-frequency pairing generally seen to require special circumstances.

To date, much of the literature on odd-frequency superconductivity has focused on its manifestation in special select circumstances, a prominent example being in the process of scattering of conventional BCS pairs into odd-frequency pairs at the interface between superconductors and other media. Examples include 
superconductor--ferromagnet interfaces~\cite{sfjunction,sfjunctionreview,linderrobinson,dibernando},
Josephson junctions~\cite{jjunction}, topological insulators~\cite{proximityti,currentdriving}, and other heterostructures. For a broader, list see Ref.~\cite{oddfreview}. Odd-frequency pairing has additionally been described as appearing in systems placed in the presence of external electric fields, where the odd-frequency is induced by Zeeman splitting~\cite{zeeman}.
Odd-frequency superconductivity has also been observed in multiband systems with inter-orbital pairing~\cite{multiband,multibandsignatures}. Odd-frequency pairing has been shown to appear ubiquitously in multiband superconductors with inter-band hybridization~\cite{multiband,oddfreview}. In this paper, we aim to demonstrate that odd-frequency pairing is also ubiquitous to the single band situation with a minimal set of degrees of freedom. 

While odd-frequency superconducting order has been observed or proposed in a variety of circumstances, the occurrence of odd-frequency superconductivity as an independent bulk phase is disputed.
It is not universally accepted that it exists as its own independent bulk state without any external influences~\cite{oddfreview,heid,fominov}.
It has been argued that the odd-frequency pairing state is inherently thermodynamically unstable~\cite{heid}. It has also been claimed that odd-frequency pairing inherently possesses an unphysical anti-Meissner effect~\cite{fominov}.
However, authors of other works have reached the opposite conclusion~\cite{belitz,solenov,kusunose}. There does not appear to be a clear consensus on this matter~\cite{oddfreview}.

To address the broader controversy regarding the intrinsic existence of odd-frequency superconductivity, we argue that a relativistic change of reference frame should not change the underlying physics.
We therefore aim to demonstrate on physical grounds that odd-frequency pairing is on equal footing with even-frequency pairing and should be considered equally as ubiquitous as implied by the symmetry table.
By ubiquity, we mean that the physical mechanism underlying odd-frequency pairing is not categorically distinct to that of conventional even-frequency pairing.

\subsection{Dynamic and relativistic considerations}
Odd-frequency superconductivity is a dynamic order with the inherent time dependence associated with the state. At equal times, the order parameter must vanish. The odd-frequency symmetry of $\Delta(t_1,t_2) = -\Delta(t_2,t_1)$ implies that $\Delta(t_1,t_2) = 0$ for $t_1 = t_2$ and therefore only makes an appearance over a finite time interval. We see now the growing list of proposals where Berezinskii state can be  realized in driven systems, such as superconductors coupled to an external driving potential~\cite{driven,multibanddriven} and Floquet engineered states~\cite{floquet}. 

We now consider the effects of a relativistic boost. Boost of electronic states in superconductors can be viewed as a dynamic process. Hence, one can ask a similar question: Will an odd-frequency state appear in the boosted superconducting state?

We start with the Galilean boost. For a homogeneous order parameter under a Galilean boost, there is no induction of odd-frequency components. A Galilean boost of an electron liquid when all electrons are in the superconducting state can be realized as a steady DC  supercurrent~\cite{currentdriving}. Assuming a fully homogeneous system with a Galilean boosted electronic liquid, the system transforms according to $\mathbf{p} \to \mathbf{p+q} $ and $\epsilon(\mathbf{p}) \to \epsilon( \mathbf{p} + \mathbf{q})$, and the gap function of the condensate as $\Delta \to \Delta \exp(\i \mathbf{q} \cdot \mathbf {r})$. We therefore find no induction of an odd-frequency state. In the inhomogeneous case, on the other hand, an odd-frequency component can be induced by a Galilean boost. Additionally, as will be shown below, in contrast to the Galilean boost, a Lorentz boost of a homogeneous order parameter induces an odd-frequency component.

The purpose of this paper is to demonstrate that odd-frequency pairing can arise intrinsically in a dynamical manner within superconducting systems from a relativistic perspective\footnote{Throughout the remainder of this paper we use the term ``relativistic'' to refer to Einsteinian relativity, and ``non-relativistic'' to refer to its $v \ll c$ limit, as well as to Galilean relativity (which forms the basis of conventional condensed matter theory).}.

As discussed previously, even- and odd-frequency pairings appear with equal weight in the symmetry classification of Table~\ref{tab:Tab1}.
On the other hand strong coupling, needed to produce odd-frequency states, often makes the odd-frequency state a subleading pairing channel. Dynamic induction of an odd-frequency channel as a result of Lorentz boost is another opportunity to induce the odd-frequency correlations.

To illustrate heuristically the influence a Lorentz boost can have on superconducting states, we can consider as a particular example a $p$-wave equal spin superconductor with a pairing of the form $\Delta(k) = \Delta_0 \sin(k)$. Under a Lorentz transformation, $\sin(k)$ transforms as
\begin{equation}
\begin{aligned}
    \sin(k) \mapsto\
    &\sin(\gamma k + \beta\gamma \omega)
    \\
    &= \underbrace{\sin(\gamma k)\cos(\beta\gamma\omega)}_{\text{even-}\omega} + \underbrace{\sin(\beta\gamma\omega) \cos(\gamma k)}_{\text{odd-}\omega} \,,
    \label{eq:heuristic}
\end{aligned}
\end{equation}
which we see manifestly generates an odd-$\omega$ term.

\begin{figure}[htp!]
\includegraphics[scale=1]{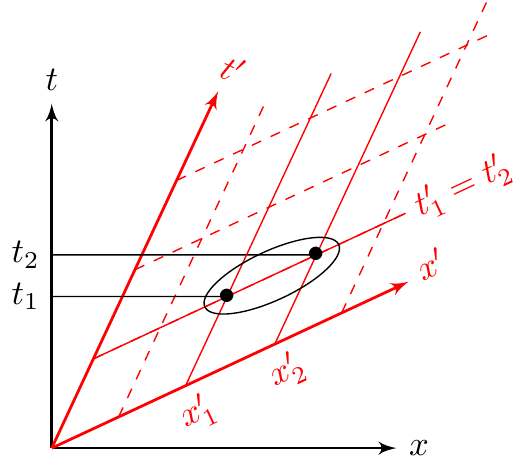}
\caption{A Cooper pair in frame $\mathcal{S}'$ (red) as viewed from frame $\mathcal{S}$ (black). The points denote the constituent fermions of the Cooper pair. Planes of equal time in $\mathcal{S}'$ are those parallel to the $x'$ axis. The time coordinates of the Cooper pair are coincident in the $\mathcal{S}'$ frame, $t_1' = t_2'$, but are displaced when viewed in frame $\mathcal{S}$, $t_1 \neq t_2$. The parity of the Cooper pair under $P^*$ in the $\mathcal{S}'$ frame is inherited by $T^*$ in the $\mathcal{S}$ frame. This illustrates that a temporal ordering can appear due to a relativistic change of reference frame. Since such a temporal displacement is necessary for odd-$\omega$ pairing, this illustration implies that an odd-$\omega$ pairing may emerge from a relativistic boost.
\label{fig:boostedcooperpair}}
\end{figure}
This transformation of order parameter may also be understood graphically from the following thought experiment. Consider a Cooper pair (we use the term \textit{Cooper pair} as a  general term for pairing, not necessarily indicative of BCS-like state) in a reference frame $\mathcal{S}'$ which is moving at a relativistic velocity with respect to frame $\mathcal{S}$, as shown in Fig.~\ref{fig:boostedcooperpair}. The spatial displacement of the constituent electrons comprising the Cooper pair in frame $\mathcal{S}'$ induces a temporal displacement between them in frame $\mathcal{S}$. 
This concept of a dynamically induced temporal parity in a Cooper pair by means of a boosted reference frame described by this illustration will be explored more quantitatively in Sec.~\ref{sec:boost}.

\begin{figure}[htp!]
\includegraphics[scale=1]{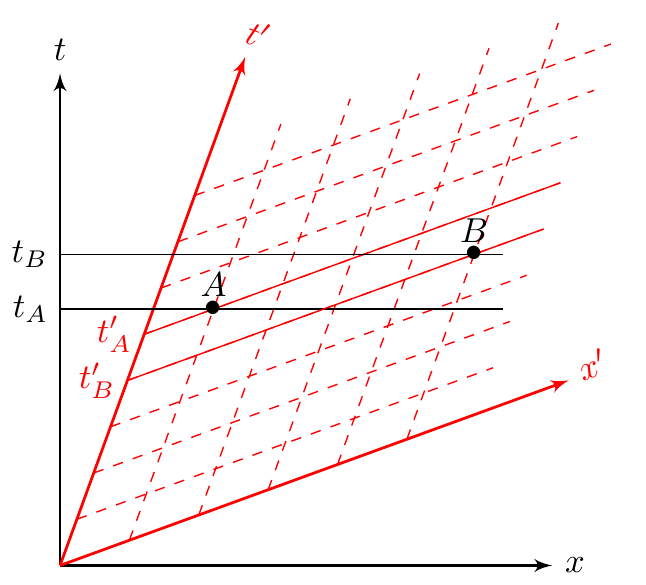}
\caption{Interchange in temporal order of space-time events by relativistic change of reference frame. Consider space-like separated events $A$ and $B$ with respect to a reference frame $\mathcal{S}$ (black) and frame $\mathcal{S}'$ (red), which is boosted with respect to $\mathcal{S}$. The lines parallel to the $x$ ($x'$) axis are planes of equal time in reference frame $\mathcal{S}$ ($\mathcal{S}'$). In frame $\mathcal{S}$, event $A$ occurs before event $B$, $t_A < t_B$. However, in the boosted frame $\mathcal{S}'$, the ordering is reversed, $t_B' < t_A'$. This shows that a relativistic change of reference frame is a manifestation of the time permutation operator $T^*$.
\label{fig:interchange}}
\end{figure}

An additional interconnection between relativity and the theory of odd-frequency superconductivity is the relation between boosts and the $T^*$ permutation operator.
The time permutation operator $T^*$ interchanges the two time coordinates of the Cooper pair.
The time permutation operation $T^*$ can manifest in the form of a relativistic frame transformation: For space-like separated spacetime events $A$ and $B$ which occur at times $t_A$ and $t_B$, respectively, and in order $t_A < t_B$ in reference frame $\mathcal{S}$, there exists a boosted reference frame $\mathcal{S}'$ such that the events occur in the order $t'_A > t'_B$. This configuration of events and frames is shown in Fig.~\ref{fig:interchange}.
We see that for such a configuration of spacetime events $A$ and $B$, changing to a boosted reference frame is a physical manifestation of the time permutation operator $T^*$.

The preceding illustrations imply that there exists some relation between relativistic transformations and odd-frequency pairing. As described above, odd-frequency pairing can be interpreted as dynamic state. In this paper, the relevant dynamics is that of a relativistic change in reference frame.
An example situation of the preceding illustrations is that of a charged particle traveling at a relativistic velocity through a particle accelerator with superconducting magnets, such as SLAC or LHC. The relevant reference frames would be the laboratory frame and the frame comoving with the accelerated particles. In the frame of the particles, the magnets of the accelerator would be observed as traveling at a relativistic velocity. The construction in this paper then describes how the superconducting magnets would appear to the accelerated particles.

The remainder of this paper is organized as follows. In Sec.~\ref{sec:theory}, we construct the theory of relativistic fermions with a pairing condensate. In Sec.~\ref{sec:boost}, we perform a Lorentz boost of an even-$\omega$ anisotropic anomalous Green function and observe the generation of odd-$\omega$ modes. We close in Sec.~\ref{sec:conclusion} with some concluding remarks.

\section{Relativistic Superconductor\label{sec:theory}}

To form a baseline for our discussion, we must first construct a theory of superconductivity which is compatible with special relativity.
This construction inherently necessitates the use of relativistic Dirac fermions. The need for Dirac fermions in chemistry and condensed matter has previously been reported in Ref.~\cite{swain}. The requirement of relativistic 4-component Dirac spinors in the description of condensed matter systems is relevant, for example, in heavy nucleon compounds, where $Z \gg 1$.
Although we are considering a relativistic scenario in the mechanical sense, with a generic superconducting object traveling at a real relativistic velocity with respect to an observer, the framework describing superconducting pairing of relativistic fermions is also relevant to heavy fermion compounds.
In comparison with the Lorentz transformation implemented in this paper, there would need to be some analog of this transformation within the material sample.
Such an effect could potentially be caused by straining the material or applying external magnetic fields~\cite{lorentzsemimetal}, but the exact implementation of such a scheme is beyond the scope of this paper, where we are simply interested in the mechanical Lorentz transformation.
The general formalism of the framework for discussing superconductivity in the relativistic regime was developed in Refs.~\cite{capellegross1,capellegross2,ohsaku1,ohsaku2} and we follow their construction in our presentation below.

The starting point of our analysis is an effective field theory defined by the action
\begin{equation}
	\begin{multlined}[c]
	S[A,\psi] = 
	\int\d^4x \left[ \overline{\psi} \left( \i \slashed{D} - m  + \frac{g_0}{2} \psi \overline{\psi} \right) \psi -\frac14 \tensor{F}{_{\mu\nu}} \tensor{F}{^{\mu\nu}} \right]
    \end{multlined}
\label{eq:action}
\end{equation}
consisting of a single flavor of relativistic Dirac fermions $\psi$ coupled to a $U(1)$ gauge field $A$ with field strength $F = \d A$ and $D = \d - \i e A$. The term quartic in the fermion field is given an attractive coupling $g_0 >0$. Units are chosen such that $\hslash = 1 = c$.
We take the standard notation $\overline{\psi}(x) \vcentcolon= \psi^\dagger(x) \gamma^0$
and use the Dirac representation of the Clifford algebra
with the convention $\gamma_5 \vcentcolon= \i \gamma^0 \gamma^1 \gamma^2 \gamma^3$ such that $(\gamma_5)^2 = +1$. Further details on conventions and explicit matrix forms of the gamma matrices are found in Appendix~\ref{sec:conventions}.
Varying the action in Eq.~\eqref{eq:action} yields the equation of motion for the fermion field as
\begin{equation}
	( \i \gamma^\mu {D}_\mu - m ) \psi(x) + g_0 (\overline{\psi}(x)^{\mathsf{T}} \psi(x)^{\mathsf{T}}) \psi(x) = 0
\label{eq:eompsi}
\end{equation}
where we rewrite $\psi(x) \overline{\psi}(x) = \overline{\psi}(x)^{\mathsf{T}} \psi(x)^{\mathsf{T}}$.
The equation of motion for the adjoint field $\overline{\psi}$ can be obtained similarly, however for the following discussion it is useful to use its transpose,
\begin{equation}
    -( \i \gamma^{\mu\mathsf{T}} {D}_\mu^{\dagger} + m ) \overline{\psi}(x)^{\mathsf{T}} + g_0 (\overline{\psi}(x) \psi(x)) \overline{\psi}(x)^{\mathsf{T}} = 0 \,.
\label{eq:eompsibar}
\end{equation}

We proceed by analyzing this system with a mean-field approximation of the interaction term.
The pairing potential obeys the
self-consistency relation
\begin{align}
	\Delta(x) &= g_0 \left\langle \psi(x)^\mathsf{T} \psi(x) \right\rangle \,,
	\label{eq:delta}
	\\
	\overline{\Delta}(x) &= g_0 \left\langle \overline{\psi}(x) \overline{\psi}(x)^{\mathsf{T}} \right\rangle \,.
	\label{eq:deltabar}
\end{align}
The pairing considered in this model is that of a generic pairing between Dirac spinors. The mechanism behind the pairing is not specified. The initial starting action defined by Eq.~\eqref{eq:action} should be considered as defining an effective field theory of an electron-positron gas and not a complete theory specifying the microscopic relativistic dynamics, e.g. we do not consider relativistic electron/positron-phonon coupling.

\subsection{Structure of the pairing term\label{sec:pairingstructure}}

For the action in Eq.~\eqref{eq:action} to be Lorentz invariant, the pairing terms must be in terms of Lorentz invariant bilinears. The general expression for the pairing potential is
\begin{equation}
	\boldsymbol{\Delta} = g_0 \left\langle \psi^{\mathsf{T}} \boldsymbol{M} \psi \right\rangle
\end{equation}
where $\boldsymbol{M}$ is a Lorentz invariant matrix which forms various combinations of pairs of spinor components.
There exist five possible bilinear classes of pairing matrices $\boldsymbol{M}$ with a total of 16 linearly independent components.
These bilinears are classified by their transformation properties under the Lorentz group as scalar, vector, bivector, pseudovector, and pseudoscalar~\cite{peskinschroeder,ohsaku1,capellegross1}.

In the ordinary non-relativistic theory of superconductivity, Cooper pairs are classified according to the symmetry of the constituent fermions' spin projections, either spin singlet or spin triplet. In the relativistic regime the spin projection is generally not a good quantum number, so it is less straightforward how to classify paired states according to spin singlet or triplet as done in the standard non-relativistic theory of superconductivity.
The pairing between the fermions is instead expressed through their  global symmetries with respect to Lorentz transformation.

To this end we express the pairing states in a configuration space in terms of time reversal $\hat{T}$ and parity $\hat{P}$ operations (not to be confused with $T^*, P^*$). In the fully relativistic scenario, we must also take into account the charge conjugation symmetry $\hat{C}$. The charge conjugation operator comes into play to describe pairing terms involving anti-particles.
In terms of the Clifford algebra basis these operators take the form of~\cite{itzyksonzuber}
\begin{subequations}
\begin{align}
	\hat{T} &= \i \gamma^1 \gamma^3 \,,
	\\
	\hat{P} &= \gamma^0 \,,
	\\
	\hat{C} &= -\i \gamma^2 \gamma^0 \,.
\end{align}
\end{subequations}
For a state of spatial momentum $\mathbf{k}$ and spin $\sigma$, $|\mathbf{k},\sigma\rangle$, the action of the time reversal and parity operators are
\begin{subequations}
\begin{align}
	| \hat{T} \, \mathbf{k},\sigma \rangle &= | -\mathbf{k},-\sigma \rangle \,,
	\\
	| \hat{P} \, \mathbf{k},\sigma \rangle &= | -\mathbf{k},\sigma \rangle \,.
\end{align}
\end{subequations}
A conventional BCS $s$-wave spin singlet pairing state is expressed as 
\begin{equation}
    |s\rangle = | \mathbf{k},\uparrow \rangle | -\mathbf{k},\downarrow \rangle - | \mathbf{k},\downarrow \rangle | -\mathbf{k},\uparrow \rangle \,.
\label{eq:swave}
\end{equation}
In terms of the symmetry operators this state may instead be expressed as
\begin{equation}
    |s\rangle = |\mathbf{k},\sigma\rangle |\hat{T}\mathbf{k},\sigma\rangle - |\hat{T}\hat{P}\mathbf{k},\sigma\rangle |\hat{P}\mathbf{k},\sigma\rangle \,.
    \tag{\ref{eq:swave}$^\prime$}
\end{equation}
For a general fermion state $|\psi\rangle$, the singlet pairing is then expressed in configuration space as 
\begin{equation}
    |s\rangle = |\psi\rangle |\hat{T}\psi\rangle - |\hat{T}\hat{P}\psi\rangle |\hat{P}\psi\rangle \,.
\end{equation}
This is the basis in which the superconducting pairing bilinears are to be constructed. The pairing matrix $\boldsymbol{M}$ is then expressed in terms of gamma matrices corresponding to the appropriate combination of symmetries $\hat{T}$, $\hat{P}$, and $\hat{C}$. Pairing terms utilizing the charge conjugation operator are pairs involving antimatter particles, either particle--anti-particle pairs or anti-particle--anti-particle pairs. These pairs will not appear in our discussion, however we note that they are needed for a complete set of relativistically invariant pairing terms.

The spin symmetry of the pairing can also be deduced from the form of the bilinears.
Bilinears satisfying $\boldsymbol{M}^{\mathsf{T}} = +\boldsymbol{M}$ correspond to triplet pairing in the non-relativistic limit and $\boldsymbol{M}^{\mathsf{T}} = -\boldsymbol{M}$ correspond to singlet pairing.
The parity of $\boldsymbol{M}$ under transposition reflects the parity of the Cooper pair under the $S$ operation in the $SP^*T^*$ classification.

A reasonable prescription for choosing the form of a pairing term in the relativistic theory is one which reduces to the standard BCS pairing term in the non-relativistic regime.
With respect to the non-relativistic limit it is useful to consider that
the four component Dirac bispinor may be decomposed into two spinors $\phi$ and $\chi$ 
\begin{equation}
	\psi = \begin{pmatrix} \phi \\ \chi \end{pmatrix}
\label{eq:psicomponents}
\end{equation}
which correspond to positive and negative energy eigenstates of the Dirac equation.

In the Dirac basis, the negative energy spinors $\chi$ are order $v/c$ compared to the positive energy spinors $\chi \sim v/c\, \phi$ \cite{feynmanqed}. Thus in the non-relativistic limit the negative energy state spinors are negligible. 
In the non-relativistic limit, the $8\times8$ matrix Gor'kov system then reduces to the standard $4\times4$ matrix system.

In the non-relativistic limit, the pairing term ${\psi}^{\mathsf{T}} \psi$ becomes $\phi^{\mathsf{T}} \phi$. The spinor $\phi$ has components
\begin{equation}
	\phi = \begin{pmatrix} \phi_{1,\uparrow} \\ \phi_{2,\downarrow} \end{pmatrix} \,.
\end{equation}
Therefore the terms of the pairing function $\boldsymbol{\Delta}$ which persist in the non-relativistic limit are those in the upper left block.
In order for the relativistic model to reduce to BCS $s$-wave singlet pairing in the non-relativistic limit, it is necessary to choose a term which is off-diagonal and antisymmetric in the upper left block of the matrix, \textit{i.e.}, a matrix whose upper left block is of the form $\begin{pmatrix} 0 & \pm 1 \\ \mp 1 & 0 \end{pmatrix}$.

The representation of a pairing between a state and its time reversed counterpart in the Clifford algebra basis is $\i \gamma_5 \gamma^0 \gamma^2$, which is a Lorentz scalar. The explicit matrix form of this term in the Dirac basis is
\begin{equation}
	- \i \gamma_5 \gamma^2 \gamma^0 = \begin{pmatrix*}[r] 0 & \phantom{-}1 & 0 & 0 \\ -1 & 0 & 0 & 0 \\ 0 & 0 & 0 & \phantom{-}1 \\ 0 & 0 & -1 & 0 \end{pmatrix*} \,.
\end{equation}
Not all possible Lorentz invariant pairing terms lead to composites which can be identified as Cooper pairs. Some of the pairing terms are pairs between positive and negative energy states, which have the interpretation of excitons in condensed matter. These pairing terms are block--off-diagonal in their matrix structure.

\subsection{The Nambu-Gor'kov system\label{sec:nambugorkov}}

Analogously to the non-relativistic formalism of superconductivity, we construct bispinor doublets in Nambu space~\cite{nambu}. Here in the relativistic case the Nambu-Dirac spinors are eight component spinors which we write as
\begin{align}
	\Psi(x) &\vcentcolon= \begin{pmatrix} \psi(x) \\ \overline{\psi}^{\mathsf{T}}(x) \end{pmatrix} \,,
	&
	\overline{\Psi}(x) &\vcentcolon= \adjvec{\ \overline{\psi}(x) & \psi^{\mathsf{T}}(x)\ } \,.
\end{align}
From these eight-component Nambu-Dirac spinors we can construct a
relativistic Gor'kov Green function \cite{gorkov}
\begin{align}
	\mathcal{G}(x,y)
	&= -\i\langle \mathcal{T} \Psi(x) \overline{\Psi}(y) \rangle
    \\
	&= \begin{pmatrix} -\i\langle \mathcal{T} \psi(x) \overline{\psi}(y) \rangle & -\i\langle \mathcal{T} \psi(x) \psi^{\mathsf{T}}(y) \rangle \\ -\i\langle \mathcal{T} \overline{\psi}^{\mathsf{T}}(x) \overline{\psi}(y) \rangle & -\i\langle T \overline{\psi}^{\mathsf{T}}(x) \psi^{\mathsf{T}}(y) \rangle\end{pmatrix}
	\\
	&=\vcentcolon \begin{pmatrix} \boldsymbol{G}(x,y) & -\i\boldsymbol{F}(x,y) \\ -\i\overline{\boldsymbol{F}}(x,y) & -\boldsymbol{G}(y,x)^{\mathsf{T}} \end{pmatrix} \,,
\label{eq:nambugreendef}
\end{align}
where $x$ and $y$ are position four-vectors and
$\mathcal{T}$ is the time ordering operator. $\boldsymbol{G}(x,y)$ and $\boldsymbol{F}(x,y)$ are the relativistic normal and anomalous Green functions respectively.
The Gor'kov Green function is an $8\times8$ matrix with the normal and anomalous Green functions $\boldsymbol{G}(x,y)$ and $\boldsymbol{F}(x,y)$ each being $4\times4$ matrices.
In the following analysis we assume the absence of external fields, $A = 0$, such that $\slashed{D} = \slashed{\partial}$.

The Gor'kov Green function obeys the equation of motion
\begin{equation}
	\begin{pmatrix} \i \slashed{\partial} - m & \boldsymbol{\Delta}(x) \\ \overline{\boldsymbol{\Delta}}(x) & \i \slashed{\partial}^{\mathsf{T}} + m \end{pmatrix} \mathcal{G}(x,y) = \boldsymbol{1} \delta(x-y)
	\label{eq:realspacegorkov}
\end{equation}
where $\slashed{\partial}^{\mathsf{T}} = (\gamma^\mu)^{\mathsf{T}} \partial_\mu$ and it is understood that $m \leftrightarrow m \boldsymbol{1}$. 
The Green function equations of motion are obtained from multiplying Eqs.~\eqref{eq:eompsi} and \eqref{eq:eompsibar} by $\overline{\psi}$ and taking the time ordered expectation value as well as utilizing the mean-field self-consistency relations Eqs.~\eqref{eq:delta} and \eqref{eq:deltabar}.
Upon Fourier transform to momentum space the Gor'kov equation Eq.~\eqref{eq:realspacegorkov} becomes
\begin{align}
	\begin{pmatrix} \slashed{k} - m & \boldsymbol{\Delta} \\ \overline{\boldsymbol{\Delta}} & \slashed{k}^{\mathsf{T}} + m \end{pmatrix} \mathcal{G}(k) &= \boldsymbol{1}
	\\
	\begin{pmatrix} \slashed{k} - m & \boldsymbol{\Delta} \\ \overline{\boldsymbol{\Delta}} & \slashed{k}^{\mathsf{T}} + m \end{pmatrix} \begin{pmatrix} \boldsymbol{G}(k) & -\i \boldsymbol{F}(k) \\ -\i \overline{\boldsymbol{F}}(k) & -\boldsymbol{G}(-k)^{\mathsf{T}} \end{pmatrix} &= \boldsymbol{1}
\end{align}
where $\slashed{k}^{\mathsf{T}} = (\gamma^\mu)^{\mathsf{T}} k_\mu$. 
This expression yields the set of coupled equations
\begin{subequations}
\begin{align}
	( \slashed{k} - m ) \boldsymbol{G}(k) - \boldsymbol{\Delta} \i\overline{\boldsymbol{F}}(k) &= \boldsymbol{1} \,,
	\label{eq:g1}\\
	-( \slashed{k}^{\mathsf{T}} + m ) \boldsymbol{G}(-k)^{\mathsf{T}} - \overline{\boldsymbol{\Delta}} \i\boldsymbol{F}(k) & = \boldsymbol{1} \,,
	\label{eq:g4}\\
	-( \slashed{k} - m ) \i\boldsymbol{F}(k) - \boldsymbol{\Delta} \boldsymbol{G}(-k)^{\mathsf{T}} &= 0 \,,
	\label{eq:g2}\\
	-( \slashed{k}^{\mathsf{T}} + m ) \i\overline{\boldsymbol{F}}(k) + \overline{\boldsymbol{\Delta}} \boldsymbol{G}(k) &= 0 \,.
	\label{eq:g3}
\end{align}
\end{subequations}
The second pair of equations, Eqs.~\eqref{eq:g2} and \eqref{eq:g3}, can be rewritten to provide closed form solutions for the first pair:
\begin{align}
	- ( \slashed{k}^{\mathsf{T}} + m ) \i \overline{\boldsymbol{F}}(k) + \overline{\boldsymbol{\Delta}} \boldsymbol{G}(k) &= 0
	\tag{\ref{eq:g3}}\\
	\overline{\boldsymbol{F}}(k) &= \i \frac{1}{\slashed{k}^{\mathsf{T}} + m} \overline{\boldsymbol{\Delta}} \boldsymbol{G}(k) \,,
	\tag{\ref{eq:g3}$^\prime$}
\end{align}
\begin{align}
	( \slashed{k} - m ) \i \boldsymbol{F}(k) + \boldsymbol{\Delta} \boldsymbol{G}(-k)^{\mathsf{T}} &= 0
	\tag{\ref{eq:g2}}\\
	\boldsymbol{G}(-k)^{\mathsf{T}} &= - \boldsymbol{\Delta}^{-1} ( \slashed{k} - m ) \i \boldsymbol{F}(k)
	\tag{\ref{eq:g2}$^\prime$} \,.
\end{align}
Inserting these expressions into the first pair of the Gor'kov equations Eqs.~\eqref{eq:g1} and \eqref{eq:g4} yields solutions for the normal and anomalous Green functions:
\begin{equation}
	\boldsymbol{G}(k) = \left[ \slashed{k} - m + \boldsymbol{\Delta} \frac{1}{\slashed{k}^{\mathsf{T}} + m} \overline{\boldsymbol{\Delta}} \right]^{-1} \,,
\end{equation}
\begin{equation}
	\i \boldsymbol{F}(k) = \left[ ( \slashed{k}^{\mathsf{T}} + m ) \boldsymbol{\Delta}^{-1} ( \slashed{k} - m) - \overline{\boldsymbol{\Delta}} \right]^{-1} \,.
\label{eq:fsolution}
\end{equation}
It is difficult to obtain explicit analytic forms for the Green functions with general pairing due to their matrix structure. However, solutions may be obtained in a straightforward manner for a given form of the pairing function $\boldsymbol{\Delta}$. A comprehensive more detailed analysis of the relativistic Gor'kov system can be found in \cite{ohsaku1,ohsaku2}. 

For gap function which transforms as the Lorentz scalar, $\boldsymbol{\Delta} = \i \Delta \gamma_5 \gamma^0 \gamma^2$, the anomalous Green function Eq.~\eqref{eq:fsolution} takes the form of
\begin{equation}
	\boldsymbol{F}(k)
	=
	\begin{pmatrix*} 0 & f(k) & 0 & 0 \\ -f^*(k)\phantom{-} & 0 & 0 & 0 \\ 0 & 0 & 0 & f(k) \\ 0 & 0 & -f^*(k)\phantom{-} & 0 \end{pmatrix*}
\label{eq:ffunction}
\end{equation}
where 
\begin{equation}
	\i f(k) = \frac{\Delta}{-k_0^2 + \mathbf{k}^2 + m^2 + |\Delta|^2} \,.
\end{equation}
$f(k)$ is a scalar function and $\boldsymbol{F}(k)$ is the full matrix function. Here we write $k_0$ for the $0^{\text{th}}$ component of the four-momentum $k$, and write $\mathbf{k}$ for the spatial components. This form of the pairing potential has previously been shown to yield a non-trivial solution to the self-consistency relation~\cite{capellegross1,capellegross2,ohsaku1,ohsaku2}.
For concreteness this function can be expressed in the explicit form of the components of $\psi(k)$ given by the Fourier transform of the definition Eq.~\eqref{eq:nambugreendef}. The general explicit form of $\boldsymbol{F}(k)$ from its definition in all its components is
\begin{align}
    \boldsymbol{F}(k)
    &=  \mathcal{T} \langle \psi(k) \psi^{\mathsf{T}}(-k) \rangle
    \\
    &=  \mathcal{T} \left\langle \begin{pmatrix}
    \phi_1 \phi_1 & \phi_1 \phi_2 & \phi_1 \chi_1 & \phi_1 \chi_2
    \\
    \phi_2 \phi_1 & \phi_2 \phi_2 & \phi_2 \chi_1 & \phi_2 \chi_2
    \\
    \chi_1 \phi_1 & \chi_1 \phi_2 & \chi_1 \chi_1 & \chi_1 \chi_2
    \\
    \phi_1 \chi_2 & \phi_2 \chi_2 & \chi_1 \chi_2 & \chi_2 \chi_2
    \end{pmatrix} \right\rangle
\label{eq:fexplicit}
\end{align}
where the elements $\phi$ and $\chi$ are the elements of $\psi$ notated in Eq.~\eqref{eq:psicomponents}. 
Comparing this expression with the particular solution Eq.~\eqref{eq:ffunction} yields the expressions
\begin{subequations}
\begin{align}
    \mathcal{T}\langle \phi_1(k) \phi_2 (-k) \rangle &= f(k)
    \\
    \mathcal{T}\langle \phi_2(k) \phi_1(-k) \rangle &= -f^*(k)
    \\
    \mathcal{T}\langle \chi_1(k) \chi_2(-k) \rangle &= f(k)
    \\
    \mathcal{T}\langle \chi_2(k) \chi_1(-k) \rangle &= -f^*(k)
\end{align}
\end{subequations}
with all other components of Eq.~\eqref{eq:fexplicit} being $0$.

In the non-relativistic limit the anomalous Green function Eq.~\eqref{eq:ffunction} reduces to the familiar $2\times2$ form of
\begin{equation}
	\boldsymbol{F}(k) = \begin{pmatrix} 0 & f(k) \\ -f^*(k) & 0 \end{pmatrix} \,,
\end{equation}
which stems from the upper left block of the full relativistic function Eq.~\eqref{eq:ffunction}.

\section{Dynamic induction of odd-frequency by Lorentz transformation\label{sec:boost}}

We now wish to consider the situation where the gap function is anisotropic, in space and/or time. Since  Lorentz boost mixes spatial and temporal degrees of freedom, an anisotropy is need in order for the Lorentz transformation to have an effect.
To demonstrate our concept of dynamically induced odd-frequency pairing, we construct an ansatz for the gap function which is valid in a particular reference frame.
We choose a frame in which the gap function takes the form of
\begin{equation}
	\Delta(k) = \frac{\Delta_0}{1 + \alpha k_0^2}
	\label{eq:gapansatz}
\end{equation}
where $\Delta_0$ is a constant and $\alpha$ parameterizes the degree of anisotropy.
This ansatz for the pairing function is explicitly even in the frequency $k_0$. At high frequency the gap function decays to zero.
In general the gap function $\Delta$ is a complex function which contains real and imaginary components.
We work in the regime where $\Re\Delta_0(0,\mathbf{k}) \approx \text{const.}$ and $\Im\Delta_0(0,\mathbf{k}) \approx 0$. The majority of literature on the analysis of pairing states stays within these assumptions. Relaxing these assumptions will not change the general conclusion about induction of odd-frequency states, but will make analysis of the main effect of Lorentz boost more tedious.

For clarity, the following discussion considers only the real part of the anomalous Green function. The imaginary part is analyzed in Appendix~\ref{sec:imagpart} for completeness.

The anomalous Green function in the frame where $\Delta$ is given by Eq.~\eqref{eq:gapansatz} takes the form of
\begin{equation}
	\i f(k) = \cfrac{\cfrac{\Delta_0}{1+\alpha k_0^2}}{-k_0^2 + \mathbf{k}^2 + m^2 + \left(\cfrac{\Delta_0}{1+\alpha k_0^2}\right)^2} \,.
\label{eq:anisof}
\end{equation}
We now wish to see the transformation of this function under a Lorentz boost.
Applying a Lorentz transformation results in a shift in the momentum four-vector by
\begin{equation}
	\begin{pmatrix} k_0 \\ {k}_\| \\ {k}_\perp \end{pmatrix}
	\mapsto
	\begin{pmatrix} \gamma & -\beta \gamma & 0 \\ -\beta \gamma & \gamma & 0 \\ 0 & 0 & \boldsymbol{1} \end{pmatrix}
	\begin{pmatrix} k_0 \\ {k}_\| \\ {k}_\perp \end{pmatrix}
\label{eq:boost}
\end{equation}
where ${k}_\|$ is comprised of the components of the momentum which are colinear in the direction of $\beta$ and ${k}_\perp$ consists of the remaining orthogonal components. The boost is parameterized by the Lorentz factor $\gamma = 1/\sqrt{1-\beta^2}$ with normalized velocity $\beta = v/c$.

Under this transformation the gap function Eq.~\eqref{eq:gapansatz} transforms as
\begin{equation}
	\Delta'(k) = \frac{\Delta_0 ( \gamma^2 k_0^2 + \beta^2 \gamma^2 k_\|^2 + 2 \beta \gamma^2 k_\| k_0 )}{1 + \alpha (\gamma^2 k_0^2 - \beta^2 \gamma^2 k_\|^2)^2} \,.
\end{equation}
The denominator is even in both $T^*$ and $P^*$. The first two terms of $\Delta'(k)$ have $T^*=+1$ and $P^* = +1$. The last term has $T^* = -1$ and $P^* = -1$ on account of being linear in both $k_0$ and $k_\|$, meaning that it is odd in frequency as well as parity. Overall the $S P^* T^*$ condition holds as $S = -1$ for spin-singlet is retained after the boost.

Applying this transformation to Eq.~\eqref{eq:anisof} results in the anomalous Green function taking the form
\begin{widetext}
\begin{equation}
	\i f'(k) = \cfrac{\Delta_0 (1-\beta^2)}{\left(1 - (1 - \alpha k_\|^2) \beta^2 - 2 \alpha k_\| k_0 \beta + \alpha k_0^2 \right) \left( k_0^2 - \mathbf{k}^2 - m^2 - \cfrac{(1-\beta^2)^2 \Delta_0^2}{(1 - (1 - \alpha k_\|^2) \beta^2 - 2 \alpha k_\| k_0 \beta + \alpha k_0^2)^2} \right)} \,.
\end{equation}
This procedure undertaken here for our analysis is analogous to a standard method of finding solutions to the Dirac equation where the solutions are first obtained in the frame where $\mathbf{k} = 0$ and then general finite-$\mathbf{k}$ solutions are obtained by boosting the rest-frame solution to an arbitrary frame where the momentum is finite~\cite{peskinschroeder}.

The characteristic properties of this transformed Green function in the boosted frame can be understood from examining the weak (low velocity) and ultra-relativistic limits.
For low boost velocity, $0<\beta\ll1$, we expand the transformed anomalous Green function in orders of $\beta$.
To first order in the boost velocity $\beta$, the anomalous Green function is
\begin{equation}
\begin{aligned}
	\i f'(k) &= \cfrac{\cfrac{\Delta_0}{1+\alpha k_0^2}}{-k_0^2 + \mathbf{k}^2 + m^2 + \left(\cfrac{\Delta_0}{1+\alpha k_0^2}\right)^2} + k_0 k_\| \beta \frac{2 \alpha \Delta_0 \left( (-k_0^2 + \mathbf{k}^2 + m^2) (1 + \alpha k_0^2)^2 - \Delta_0^2 \right)}{\left((-k_0^2 + \mathbf{k}^2 + m^2) (1 + \alpha k_0^2) + \Delta_0^2 \right)^2} + \mathcal{O}(\beta^2) \,.
\end{aligned}
\label{eq:infinitesimalboost}
\end{equation}
In the boosted frame we now analyze the system in imaginary time $\tau = -\i t$ in terms of the
Matsubara frequency $k_0 \to \i \omega_n = \pi k_{\textsc{b}}T (2n +1)$. The anomalous Green function in Matsubara frequency is defined as $\boldsymbol{F}(\tau,\mathbf{x}) = \langle \mathcal{T}_\tau \psi(\tau,\mathbf{x}) \psi^{\mathsf{T}}(0,0) \rangle$ where $\mathcal{T}_{\tau}$ is time ordering operator in imaginary time $\tau$. In Matsubara frequency, the boosted anomalous Green function is
\begin{equation}
	f'(\i\omega_n,\mathbf{k}) = \cfrac{\cfrac{\Delta_0}{1-\alpha \omega_n^2}}{\omega_n^2 + \mathbf{k}^2 + m^2 + \left(\cfrac{\Delta_0}{1-\alpha \omega_n^2}\right)^2} + \i \omega_n k_\| \beta \frac{2 \alpha \Delta_0 \left( (\omega_n^2 + \mathbf{k}^2 + m^2) (1 - \alpha \omega_n^2)^2 - \Delta_0^2 \right)}{\left((\omega_n^2 + \mathbf{k}^2 + m^2) (1 - \alpha \omega_n^2) + \Delta_0^2 \right)^2} + \mathcal{O}(\beta^2) \,.
\label{eq:fmatsubara}
\end{equation}
The transition to Matsubara frequency is introduced only at this stage after the relativistic change of reference frame in order to avoid ambiguities associated with defining the transformation properties of temperature, and hence Matsubara frequency, under Lorentz transformations \cite{temperaturemovingbody,relativistickms,movingthermodynamics,temperaturetransformations,relativistictemperature}.

In the transformed function Eq.~\eqref{eq:fmatsubara} we see that the first order relativistic correction under a Lorentz boost is exclusively odd in Matsubara frequency: The first order correction term contains terms of order $\omega_n$, $\omega_n^3$, and $\omega_n^5$.
This shows that for \textit{any} finite boost $\beta > 0$, an odd-frequency component is induced dynamically in the anomalous Green function. This is the primary result of this paper. 
The appearance of odd-frequency already at $\mathcal{O}(\beta^1)$ reinforces the notion that odd-frequency pairing is a ubiquitous state. For comparison, spin-orbit coupling, which is known to occur in a wide variety of materials, is an effect which occurs at $\mathcal{O}(\beta^2)$~\cite{spinorbitreview,feynmanqed}.

The magnitude of the odd-frequency contribution to the anomalous propagator is proportional to the magnitude of the boost, i.e. $v/c$.

Next, we wish to examine the behavior of the anomalous Green function under a boost in the ultra-relativistic limit where $\beta \sim 1^-$. Performing an expansion of $f'(k)$ around $(1-\beta)$ we find that
\begin{equation}
	f'(\i\omega_n,\mathbf{k})
	=
	\frac{2 \Delta_0 (1-\beta)}{\alpha (\i\omega_n - k_\|)^2 (-\omega_n^2 - \mathbf{k}^2 - m^2)} 
	- \frac{\Delta_0 ( 4 - 3 \alpha k_\|^2 + 2 \alpha k_\| \i \omega_n - \alpha \omega_n^2 ) (1-\beta)^2}{\alpha^2 (\i\omega_n - k_\|)^4 (-\omega_n^2 - \mathbf{k}^2 - m^2)}
	+ \mathcal{O}((1-\beta)^3) \,,
\label{eq:ultralimit}
\end{equation}
\end{widetext}
where the first order term can be rewritten as
\begin{equation}
\begin{multlined}
	\frac{2 \Delta_0 (1-\beta)}{\alpha (\i\omega_n - k_\|)^2 (-\omega_n^2 - \mathbf{k}^2 - m^2)}
	\\=
	\frac{2 \Delta_0 (-\omega_n^2 + k_\|^2 + 2 k_\| \i\omega_n) (1-\beta)}{\alpha (-\omega_n^2 + k_\|^2)^2 (-\omega_n^2 - \mathbf{k}^2 - m^2)} \,.
\end{multlined}
\end{equation}
Here we see that this term contains terms which are both even- and odd-powers of Matsubara frequency. Therefore, in the ultra-relativistic limit the superconducting order parameter is neither exclusively even- or odd-frequency for a given order of $(1-\beta)$. Terms of both symmetries must be present.
The second order term in the ultra-relativistic expansion shows the same characteristic, indicating that this feature is generic in the ultra-relativistic limit.
In the extreme limit of $\beta = 1$, the anomalous Green function of the spin singlet sector vanishes.

Note that terms which are odd under $T^*$ are always also odd under $P^*$ with the combination $T^* P^* = +1$. This reflects that a Lorentz transformation does not violate the $SP^*T^* = -1$ restriction as the spin degree of freedom remains in the singlet sector, $S=-1$.

As shown from the above, the anomalous Green function contains both even-$\omega$ and odd-$\omega$ components for any finite boost. The odd-$\omega$ components appear on equal footing with the even-$\omega$ terms.

\section{Conclusion\label{sec:conclusion}}

In this paper we analyzed the consequences of viewing a conventional even-frequency superconductor from a relativistically boosted observer.
We therefore conclude that the pairing between electrons in odd-frequency Cooper pairs is physically equivalent to the pairing between conventional even-frequency Cooper pairs since relativistic transformations should not change the underlying physics. The concept of odd-frequency pairing should then be thought of as ubiquitous as even-frequency pairing.
An analogy here is with electric and magnetic fields. ubiquity of magnetic fields in comparison to electric fields.
Macroscopic electric and magnetic fields are physically distinct with non-identical properties, however their interpretation as components of the electromagnetic field imply that one is no more ubiquitous than the other. Similarly, we can say that even-frequency pairing in superconductivity is no more ubiquitous than odd-frequency pairing.

We note that the notion of a relativistically boosted Cooper pair has previously been studied with respect to the entanglement structure of the constituent fermions~\cite{vlatko1,vlatko2}. These studies concluded that an isotropic $s$-wave spin singlet Cooper pair would acquire spin triplet components under a Lorentz boost due to Wigner rotation~\cite{wigner}. This result appears to differ from the relativistic theory of superconductivity constructed above based on~\cite{capellegross1,capellegross2,ohsaku1,ohsaku2}, as for example the spin singlet pairing correlation function Eq.~\eqref{eq:ffunction} is invariant under Lorentz transformations and does not acquire triplet components (recall that spin triplet pairing manifests as even parity under matrix transposition). 
This difference in the result is presumably related to the fact that~\cite{vlatko1,vlatko2} label their states with the non-relativistic scheme of different quantum numbers for orbital- and spin-angular momentum separately, despite these not being good quantum numbers relativistically.
%
With respect to the conclusions of~\cite{vlatko1,vlatko2} where a boosted spin singlet transforms into a pair with singlet and triplet components, our work presented above can be said to occur in the spin singlet projection of the transformed pair, and thus our conclusions regarding the emergence of an odd-frequency channel are valid regardless.

The role of relativity with respect to odd-frequency pairing has ramifications to other systems beyond the simple condensed matter model here. It is suspected that there exists a region of the quantum chromodynamic (QCD) phase diagram in which quarks form pairs in a condensate, which is known as color superconductivity~\cite{qcdsc}. As we showed in this work, there exists a relationship between relativity and odd-frequency pairing, which implies that odd-frequency color Cooper pairs in principle should appear in the color superconducting state.

In summary, we presented a calculation demonstrating the dynamical emergence of an odd-$\omega$ superconductivity channel as a result of a Lorentzian change in reference frame.
We first constructed a theory of superconductivity which is relativistically invariant for constant gap function $\Delta$. We then supposed a temporally anisotropic form of the gap function even in frequency in a particular reference frame. The anomalous Green function for this temporally anisotropic gap function was then boosted into a new frame, where we found that an odd-$\omega$ channel appears. 
The result of this calculation is that it provides an existence proof for the notion that odd-$\omega$ superconductivity is a valid channel inherent to superconductors which appears ubiquitously.
Furthermore, this work implies that in order for superconductivity to be compatible with the fundamental symmetry of special relativity, odd-frequency should generally be considered.

\begin{acknowledgments}
We are grateful to A. Black-Schaffer and M. Geilhufe for useful discussions. We acknowledge support from the Knut and Alice Wallenberg Foundation KAW 2019.0068,  European Research Council under the European Union Seventh Framework ERS-2018-SYG 810451 HERO and  the University of Connecticut.
\end{acknowledgments}

\appendix
\section{Conventions\label{sec:conventions}}
In this appendix are some conventions of the gamma matrices used in the main text.
\begin{align}
	\gamma^0 &= \begin{pmatrix*}[r] \boldsymbol{1} & \boldsymbol{0} \\ \boldsymbol{0} & -\boldsymbol{1} \end{pmatrix*}
	&
	\gamma^{j} &= \begin{pmatrix} \boldsymbol{0} & \boldsymbol{\sigma}^j \\ -\boldsymbol{\sigma}^j & \boldsymbol{0} \end{pmatrix}
\end{align}
\begin{align}
	\gamma_5 &\vcentcolon= \i \gamma^0 \gamma^1 \gamma^2 \gamma^3
	= \begin{pmatrix} \boldsymbol{0} & \boldsymbol{1} \\ \boldsymbol{1} & \boldsymbol{0} \end{pmatrix}
\end{align}
\begin{align}
	\{ \gamma^\mu , \gamma^\nu \} &= 2 \eta^{\mu\nu} & \boldsymbol{\eta} &= \text{diag}[+,-,-,-] 
\end{align}
\begin{align}
	(\gamma^0)^\dagger &= \gamma^0 & (\gamma^j)^\dagger &= -\gamma^j
\end{align}

\section{Imaginary part of $f(k)$\label{sec:imagpart}}
In this appendix we analyze the imaginary part of the anomalous Green function Eq.~\eqref{eq:anisof} for the form of the gap function Eq.~\eqref{eq:gapansatz} under application of the boost Eq.~\eqref{eq:boost}. For an even-$\omega$ gap function, it holds that $\Re\Delta(\omega)$ is even in $\omega$ and $\Im\Delta(\omega)$ is odd in $\omega$. Likewise, for an odd-$\omega$ gap function, $\Re\Delta(\omega)$ is odd in $\omega$ and $\Im\Delta(\omega)$ is even in $\omega$. For completeness, it is necessary to check that the Lorentz transformation of the anomalous Green function as performed in the main text satisfies these conditions.

Taking the anomalous Green function Eq.~\eqref{eq:anisof}, we apply the prescription of $k_0 \mapsto k_0 + \i\delta$ for $0 < \delta \ll 1$ to obtain
\begin{equation}
    f(k) = \cfrac{\cfrac{\Delta_0}{1 + \alpha (k_0+\i\delta)^2}}{-(k_0+\i\delta)^2 + \mathbf{k}^2 + m^2 + \left(\cfrac{\Delta_0}{1 + \alpha (k_0+\i\delta)^2}\right)^2} \,.
\end{equation}
We obtain the imaginary part by expanding the expression to first order in $\delta$.
To first order in $\delta$, $\mathcal{O}(\delta)$, the imaginary part of the anomalous Green function is
\begin{widetext}
\begin{equation}
    \Im f(k) \Big\vert_{\mathcal{O}(\delta)} = - 2 k_0 \Delta_0 \cfrac{(1 - \alpha(-2k_0^2 + \mathbf{k}^2 + m^2)) (1+\alpha k_0^2)^2 + \alpha \Delta_0^2}{\left((-k_0^2 + \mathbf{k}^2 + m^2) (1+\alpha k_0^2)^2 + \Delta_0^2\right)^2} \,.
\end{equation}
To this function we apply the Lorentz boost from Eq.~\eqref{eq:boost}.
The weak asymptotic limit of $0<\beta\ll1$ takes the form
\begin{equation}
    \Im f'(k) \Big\vert_{\mathcal{O}(\delta)} = 
    \begin{multlined}[t]
    2 k_\| \Delta_0 \beta \left[ \cfrac{\left( (-k_0^2 + \mathbf{k}^2 + m^2) (1+\alpha k_0^2)^3 \left(1 - \alpha( \mathbf{k}^2 + m^2) + \alpha (1 + 3 \alpha(\mathbf{k}^2+m^2)\right) k_0^2 - 4 \alpha^2 k_0^4\right)}{\left((-k_0^2 + \mathbf{k}^2 + m^2) (1+\alpha k_0^2)^2 + \Delta_0^2\right)^3} \right.
    \\
    - 
    \left.
    \cfrac{(1+\alpha k_0^2)\left(1 + \alpha k_0^2 (8 - 12 \alpha (\mathbf{k}^2+m^2) + 19 \alpha k_0^2)\right) \Delta_0^2 - \alpha \Delta_0^4}{\left((-k_0^2 + \mathbf{k}^2 + m^2) (1+\alpha k_0^2)^2 + \Delta_0^2\right)^3} \right]
    + \mathcal{O}(\beta^2) \,.
    \end{multlined}
\label{eq:imagboostedf}
\end{equation}
By inspection this function is even in $k_0$ and odd in $k_\|$, which is the opposite parity of the first order term in the real part in Eq.~\eqref{eq:infinitesimalboost} as expected.

In the ultra-relativistic asymptotic limit $\beta \sim 1^-$, the imaginary part to first order in $(1-\beta)$ reads as
\begin{equation}
    \Im f'(k) \Big\vert_{\mathcal{O}(\delta)} = 
    \begin{multlined}[t]
    \cfrac{2\sqrt{2} \Delta_0 (k_0 + k_\|) \sqrt{1-\beta}}{\alpha (-k_0^2 + k_\|^2) (-k_0^2 + \mathbf{k}^2 + m^2)^2} + \mathcal{O}((1-\beta)^{3/2}) \,.
    \end{multlined}
\end{equation}
\end{widetext}
As in the real part of Eq.~\eqref{eq:ultralimit}, this function contains pieces which are both even and odd in $k_0$. This matches the result that the even- and odd-frequency components get mixed together to same order in the boost parameter in the ultra-relativistic limit.

For odd-frequency pairing, the anomalous spectrum, which is given by the imaginary part of the anomalous Green function, is \textit{even}-frequency. 
The anomalous spectral function is obtained from the imaginary part of the retarded anomalous Green function as\footnote{As the boosted anomalous propagator is odd in $k_\|$, the integration domain is only over the half line on this component, $k_\| \in [0,\infty)$, otherwise $\mathcal{B}^{\text{even}}(\omega)$ would trivially be zero.}
\begin{equation}
    \mathcal{B}(\omega) = -\frac1\pi \Im \int \d^3\mathbf{k} f(\omega+\i0^+,\mathbf{k}) \,.
\end{equation}
The symmetry of the spectral function can be observed from its decomposition into even and odd parts as $\mathcal{B}^{\text{even}}(\omega) = \frac12 \left[ \mathcal{B}(\omega) + \mathcal{B}(-\omega) \right]$ and $\mathcal{B}^{\text{odd}}(\omega) = \frac12 \left[ \mathcal{B}(\omega) - \mathcal{B}(-\omega) \right]$.

\begin{figure}
\centering
\includegraphics[scale=1]{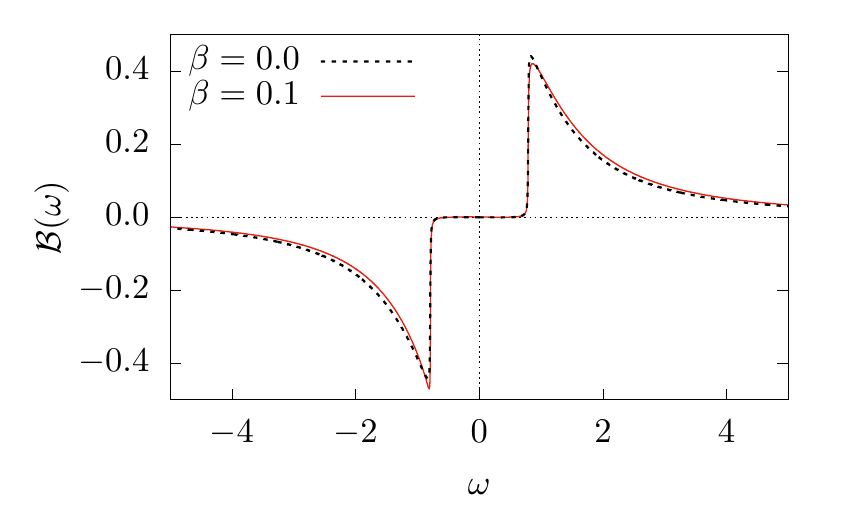}
\caption{The anomalous spectrum $\mathcal{B}(\omega)$ in arbitrary units with the parameters $\Delta_0 = 1.0$, $m = 0.5$, and $\alpha = 1.0$, without boost ($\beta = 0$, black dashed) and with finite boost ($\beta = 0.1$, red solid). Odd-$\omega$ pairing corresponds to even-$\omega$ symmetry in the imaginary part of $f(k)$ plotted here. Under a finite boost $\mathcal{B}(\omega)$ is a function which contains both even- and odd-$\omega$ contributions. The even-$\omega$ contribution is small in magnitude compared to the odd-$\omega$ contribution, \textit{cf}. Fig.~\ref{fig:boostedimf}.
\label{fig:boostedimftotal}}
\end{figure}
\begin{figure}
\centering
\includegraphics[scale=1]{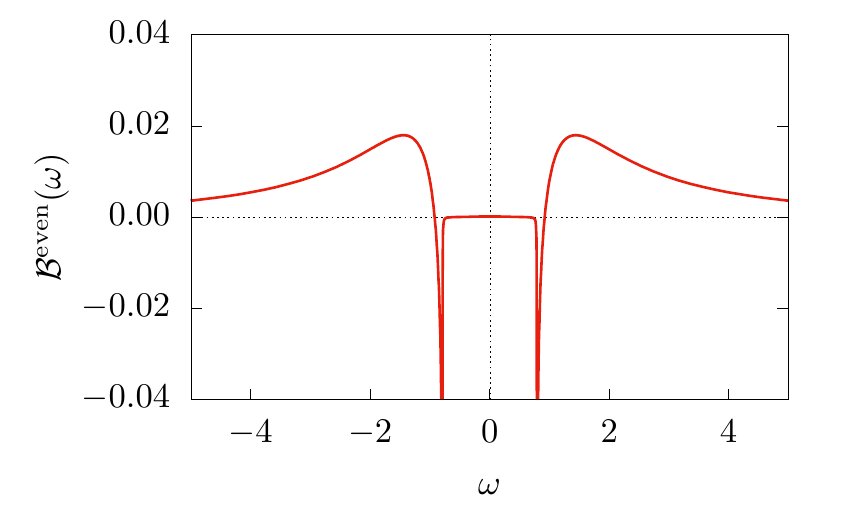}
\caption{The even-$\omega$ part $\mathcal{B}^{\text{even}}(\omega) = \frac12 \left[ \mathcal{B}(\omega) + \mathcal{B}(-\omega) \right]$ of the anomalous spectrum for finite boost, $\beta = 0.1$, plotted in arbitrary units with the parameters $\Delta_0 = 1.0$, $m = 0.5$, and $\alpha = 1.0$. The even-$\omega$ in the imaginary part of the spectrum plotted here corresponds to odd-frequency pairing. In the unboosted case $\mathcal{B}^{\text{even}}(\omega)=0$ $\forall \omega$. Note that the even contribution is an order of magnitude smaller in comparison to the total anomalous spectral function plotted in Fig.~\ref{fig:boostedimftotal}.
\label{fig:boostedimf}}
\end{figure}

In the absence of a relativistic boost, there is no even-$\omega$ contribution to $\mathcal{B}(\omega)$. For a finite boost, $\mathcal{B}(\omega)$ acquires an even-frequency contribution.
The anomalous spectral function for a small finite boost, $\beta = 0.1$, in comparison to the unboosted spectrum is plotted in Fig.~\ref{fig:boostedimftotal} with the parameters $\Delta_0 = 1.0$, $m = 0.5$, and $\alpha = 1.0$. The spectrum is computed numerically with $\delta=0.005$. 
We compute these integrals numerically with finite bandwidths. The integrals were computed for a variety of bandwidths and were found to produce the same solution. We therefore conclude that the integrals are convergent.

The even contribution $\mathcal{B}^{\text{even}}(\omega)$ is plotted in Fig.~\ref{fig:boostedimf}. While the overall spectrum appears odd, there is a slight even component which emerges. As shown by Fig.~\ref{fig:boostedimf}, the even contribution is an order of magnitude smaller than the total function. Graphically we see that $\mathcal{B}^{\text{even}}(\omega) \neq 0$ for $\beta > 0$, indicating that a finite boost induces an even-frequency component in the imaginary part of the anomalous Green function, as would be expected in the case of odd-frequency pairing.

These computations show that the imaginary part of the anomalous Green function behaves as expected for the generation of odd-frequency components in the real part.

\bibliography{oddfreferences}

\end{document}